\documentclass[nojss]{jss}

\usepackage{thumbpdf,lmodern}

\usepackage{framed}
\usepackage{bbold}
\usepackage{bm}
\usepackage{amsmath}
\usepackage{algorithm}
\usepackage{algpseudocode}
\usepackage{enumitem}
\usepackage{mdwlist}
\usepackage{xcolor}
\usepackage{multirow}

\newcommand{\fct}[1]{\code{#1()}}

\author{Charlotte Baey\\Univ. Lille, CNRS, UMR 8524\\ Laboratoire Paul Painlevé, F-59000 Lille, France
   \And Estelle Kuhn\\MaIAGE, Université Paris-Saclay, \\INRAE, France} 
\Plainauthor{Charlotte Baey, Estelle Kuhn}

\title{\pkg{varTestnlme}: an \proglang{R} package for Variance Components Testing in Linear and Nonlinear Mixed-effects Models}
\Plaintitle{an R package for Variance Components Testing in Generalized Linear and Nonlinear Mixed-effects Models}
\Shorttitle{\pkg{varTestnlme}: Variance Components Testing in Mixed-effects Models}

\Abstract{
The issue of variance components testing arises naturally when building mixed-effects models, to decide which effects should be modeled as fixed or random. While tests for fixed effects are available in \proglang{R} for models fitted with \pkg{lme4}, tools are missing when it comes to random effects.
  The \pkg{varTestnlme} package for \proglang{R} aims at filling this gap. It allows to test whether any subset of the variances and covariances corresponding to any subset of the random effects, are equal to zero  using asymptotic property of the likelihood ratio test statistic. It also offers the possibility to test simultaneously  for fixed  effects and variance components. It can be used for linear, generalized linear or nonlinear mixed-effects models fitted via \pkg{lme4}, \pkg{nlme} or \pkg{saemix}.  Theoretical properties of the used likelihood ratio test  are recalled, numerical methods used to implement the test procedure are detailed and  examples based on different real datasets using different mixed models are provided.
}

\Plainkeywords{generalized mixed-effects models, nonlinear mixed-effects models, variance components, likelihood ratio test, random effects, R}

\Address{
  Charlotte Baey\\
  Laboratoire Paul Painlev\'e\\
  Universit\'e de Lille\\
  Cit\'e Scientifique
  59655, VIlleneuve d'Ascq, France\\
  E-mail: \email{charlotte.baey@univ-lille.fr}\\
  URL: \url{http://math.univ-lille1.fr/~baey/}\\
  
  Estelle Kuhn\\
  INRAE unit\'e MaIAGE\\
  Domaine de Vilvert\\
  78352 Jouy-en-Josas, France\\
  E-mail: \email{estelle.kuhn@inrae.fr}\\
  URL: \url{http://genome.jouy.inra.fr/~ekuhn/}
}

\begin{document}



\section{Introduction}\label{sec:intro}

Mixed-effects models are widely used in many fields of applications, ranging from agronomy, social science, medical science, or biology. One can distinguish three main types of mixed models: linear mixed models (LMMs), generalized linear mixed models (GLMMs) and nonlinear mixed models (NLMMs).
Several packages and softwares are available to fit these types of model, for example the \proglang{SAS} procedures \code{MIXED} and \code{NLMIXED} for linear, generalized linear and nonlinear mixed models, or the dedicated softwares \proglang{NONMEN} and \proglang{Monolix} for nonlinear mixed models, with an emphasis on pharmacodynamic and pharmacokinetic models for the former. In \proglang{R}, three main packages are available to deal with these models: \pkg{lme4} \citep{lme4}, \pkg{nlme} \citep{nlme} and \pkg{saemix} \citep{saemix}. If they can all deal with the three types of mixed models, the way they handle nonlinear mixed models differs. Both \pkg{lme4} and \pkg{nlme} use an approximation of the likelihood function based on a linearization of the model, while \pkg{saemix} uses an EM-type algorithm to derive the maximum likelihood estimate, without resorting to any approximation or linearization of the model.

When building a mixed-effects model, one is facing the  delicate choice of which effects should be modeled as random. This issue has been studied both from  theoretical and  practical point of views. 

On the theoretical side, two main approaches have been explored. The first one follows the idea of variable selection. \cite{Chen03} focused on linear mixed models and proposed a Bayesian formulation using mixture priors with a point mass at zero for the random effects. More recently, \cite{Ibra11,Groll14} used a penalized likelihood approach to simultaneously select fixed and random effects. Selection criteria adapted to the context of mixed-effects models were also suggested by \cite{Vai05,Gurka06} and \cite{Del14}. The second approach concerns  hypothesis testing based on the variance components, with the seminal work of \cite{stramlee94,stramlee95} in linear mixed models (see also the review by \cite{Mol07}). Recently, \cite{bck19} have exhibited the exact limiting distribution of the likelihood ratio test (LRT) statistic, for testing that the variances of any subset of the random effects are equal to zero. 

On the practical side, several of the aforementioned methods have been implemented e.g., in \proglang{R}. Among others, we can cite for example the \pkg{glmmLasso} \citep{glmmLasso} package, computing an $\ell_1$-penalized maximum likelihood estimator in the context of generalized linear mixed models, or the \pkg{cAIC4} \citep{caic4} package, implementing the conditional AIC criterion. 
As far as hypothesis testing is concerned, most of the available tools are designed for linear or generalized linear mixed-effects models, and are mostly adapted to outputs from \pkg{lme4} package. The \pkg{lmerTest} package \citep{lmertest} provides enhanced versions of functions \fct{anova} and \fct{summary} of the \pkg{lme4} package by computing $p$-values for the corresponding tests for fixed effects. It also implements both the Satterthwaite's and the Kenward-Roger's approximations to correct $p$-values in unbalanced or small sample size situations. It also provides a \fct{step} function, which simplify the random effects structure of the model using a step-down approach. Each step is based on the computation of $p$-values from the LRT for testing that the variance of each random effect is equal to zero. However, the asymptotic distribution used to compute these $p$-values is a chi-square one, with the degree of freedom computed as the difference between the number of parameters under the alternative hypothesis and the number of parameters under the null hypothesis.
In the context of mean components testing in linear and generalized linear models, the \pkg{pbkrtest} package \citep{pbkrtest} provides two alternatives to the LRT in small sample size situations, based either on the Kenward-Roger approximation or on parametric bootstrap. The two aforementioned packages can only be used with models fitted with the \pkg{lme4} package. 
The  \pkg{RLRSim} package implements the tests proposed by \cite{Crai04a}, based on the exact finite sample distribution of the LRT statistic. It can be used with a larger variety of packages and can handle not only models fitted from packages \pkg{lme4} and \pkg{nlme}, but also from package \pkg{gamm4} and from function \code{gamm} of package \pkg{mgcv}.
However, only linear mixed models with one single random effect are currently covered by this package.
In \proglang{SAS}, the option \code{COVTEST} can be used for linear mixed models in \code{PROC MIXED} and provides Wald tests for the variance components. More recently, a \proglang{SAS} macro \code{\%COVTEST} performing a bootstrap parametric test has been proposed for linear mixed models.

To the best of our knowledge, there is no software or \proglang{R} package implementing the LRT for  variance components in all types of mixed-effects models, and for an arbitrary subset  of variance components corresponding to any subset of the random effects.

In this paper, we present the \pkg{varTestnlme} \proglang{R} package dedicated  to test that the variances of any subset of the random effects are equal to zero in LMMs, GLMMs and NLMMs, using models that were fitted using either \pkg{lme4}, \pkg{nlme} or \pkg{saemix}. Moreover it is  possible to test simultaneously mean and variance components.
The \pkg{varTestnlme} package takes as inputs the two competitor models, fitted with the same package, and returns the $p$-value based on the asymptotic distribution of the LRT statistic associated with the two nested models corresponding to the null and alternative hypotheses. 
The theory behind this test has been established in \cite{bck19}.The purpose of the current paper is to provide hints for practical implementation and a complete description of the different numerical methods used for the computation of each quantity involved. In particular, we detailed the  algorithms used to sample from the limiting chi-bar-square distribution and to estimate the weights of the different chi-square distributions appearing in this mixture. We also extend results of \cite{bck19}, where only the cases of a diagonal or full covariance matrix for the random effects were exemplified, to the general case of a block-diagonal covariance matrix, covering a large  range of covariance structures.

Mixed-effects models and the LRT procedure are recalled in Section \ref{sec:modelandtest}. The implementation of the method is described in Section \ref{sec:implementation}. Section \ref{sec:illustrations} is devoted to illustrations and examples of how to use the package. Finally, current limitations and possible extensions are discussed in Section \ref{sec:summary}.

\section{Mixed-effects models and likelihood ratio test procedure} \label{sec:modelandtest}

\subsection{Description of the mixed-effects models}
We consider the following nonlinear mixed-effects model (\cite{Dav95} p98, \cite{Pin00} p306, \cite{Lavielle2014} p24):
\begin{equation}\label{eq:intravar}
y_{i} = g(\varphi_i, x_{i}) + \varepsilon_{i},
\end{equation}
where $y_{i}$ denotes the vector of observations of the $i$-th individual of size $J$, $1 \leq i \leq n$, $g$ a nonlinear function, $\varphi_i$  the vector of individual parameters 
 of individual $i$,  $x_{i}$ the vector of covariates, and $\varepsilon_i$ the random error term. The vectors of individual parameters $(\varphi_i)_{1 \leq i \leq n}$ are modeled as:
\begin{equation}\label{eq:inter_var}
\varphi_i =U_i \beta + V_i b_i \ \ ,  \ 1 \leq i \leq n,
\end{equation}
where $\beta$ is the vector of fixed effects  taking values in $\mathbb{R}^b$, $U_i$ and $V_i$ are covariates matrices of individual $i$ of sizes $p\times b$ and $p\times p$ respectively, $b_i$ is the centered vector of Gaussian random effects  with covariance matrix $\Gamma$ of size $p \times p$. 
The random vectors $(b_i)_{ 1 \leq i \leq n}$ are assumed to be independent. 
The vectors $(\varepsilon_{i})_{ 1 \leq i \leq n}$ are assumed to be independent and identically distributed centered Gaussian vectors with covariance matrix  $\Sigma$. Finally the sequences $(\varepsilon_{i})$ and $(b_i)$ are assumed  mutually independent.

Let $\theta=(\beta, \Gamma,\Sigma)$ denotes the parameters vector of the model taking value in the set $\Theta= \{ \theta \in \mathbb{R}^q \mid  \ \beta \in \mathbb{R}^b,  \Gamma \in \mathbb{S}^{p}_+, \Sigma \in \mathbb{S}^J_+\}$ where 
$\mathbb{S}^{a}_+$ denotes the set of positive definite matrices of size $a$.

\subsection{Testing variance components}
We consider general test hypotheses of the following form, to test for the nullity of $r$ variances among $p$:
\begin{equation}\label{eq:hypotheses}
H_0 : \theta \in \Theta_0 \quad \text{against} \quad H_1 : \theta \in \Theta,
\end{equation}
where $\Theta_0 \subset \Theta$. 
Up to permutations of rows and columns of the covariance matrix $\Gamma$, we can assume that we are testing the nullity of the last $r$ variances. We write $\Gamma$ in blocks as follows:
\begin{equation*}
\Gamma = \left( \begin{array}{c|c}
\Gamma_1 &  \Gamma_{12}^\top \\
\hline
\Gamma_{12} & \Gamma_2
\end{array}
\right),
\end{equation*}
with $\Gamma_1$ a $(p-r) \times (p-r)$ matrix, $\Gamma_2$ a $r \times r$ matrix, $\Gamma_{12}$ a $r \times (p-r)$ matrix and where $A^\top$ denotes the transposition of matrix $A$, for any matrix $A$.

Thus we have:
\begin{align*}
	\Theta_0 & = \{\theta \in  \mathbb{R}^q \mid \beta \in \mathbb{R}^b, \Gamma_1 \in \mathbb{S}_+^{p-r}, \Gamma_{12} = 0, \Gamma_2 = 0, \Sigma \in \mathbb{S}^{J}_+ \} \\
	\Theta & = \{\theta \in  \mathbb{R}^q \mid \beta \in \mathbb{R}^b, \Gamma \in \mathbb{S}_+^{p}, \Sigma \in \mathbb{S}^{J}_+ \}.
\end{align*}


The likelihood ratio test (LRT) statistic is then defined by:
\[
LRT_n  = -2 \ \log \left( \frac{\sup_{\theta \in \Theta_0}  L (y_1, \dots, y_{n} ;\theta)}{\sup_{\theta \in \Theta}L (y_1, \dots, y_{n} ; \theta)} \right),
\]
where $L (y_1, \dots, y_{n} ;\theta)$ is the likelihood function.

As shown in \cite{bck19}, the limiting distribution of the LRT statistic associated with the two hypotheses $H_0$ and $H_1$ defined  in \eqref{eq:hypotheses} is:
\begin{gather}\label{eq:lrtdist}
	LRT_n \xrightarrow[n  \rightarrow \infty]{} \bar{\chi}^2(I^{-1}_*,T(\Theta,\theta^*)\cap T(\Theta_0,\theta^*)^{\perp}),
\end{gather}
with $I_*$ is the Fisher Information Matrix, $T(\mathcal{A},\theta)$ is the tangent cone to the space $\mathcal{A}$ at point $\theta$, and where $\bar{\chi}^2(V,\mathcal{C})$ denotes the chi-bar-square distribution  parametrized by the positive definite matrix $V$ and by the cone $\mathcal{C}$. The chi-bar-square distribution is a mixture of  chi-bar distributions, where the degrees of freedom and the weights involved in the mixture depend on the matrix $V$ and on the cone $\mathcal{C}$.
More details about the computation of the parameters of the chi-bar square distribution are given in Section \ref{sec:implementation}.

\subsection{Testing simultaneously fixed effects and variance components}

Results of the previous section can be  extended to the case where one is testing simultaneously that a subset of the mean parameters and a subset of the covariance parameters are null. In this case, the null hypothesis and subsequently the parameter space $\Theta_0$ are slightly modified as follows: if we denote by $r_f$ the number of fixed effects which are tested equal to 0, then we need to replace ``$\beta \in \mathbb{R}^b$'' by ``$\forall k =1,\dots, r_f, \beta_k = 0, \text{and } \forall k = r_f+1, \dots, b, \beta_k \in \mathbb{R}$'' in the definition of $\Theta_0$ (see \eqref{eq:hyp0}).
The theoretical result also holds in this test setting, leading to a chi-bar square distribution.

\section{Implementation} \label{sec:implementation}

\subsection{Parameters of the chi-bar-square distribution}\label{cone}
The parameters of the chi-bar-square distribution appearing in Equation \eqref{eq:lrtdist} as the limiting distribution of the LRT statistic are the Fisher Information Matrix and the tangent cone $T(\Theta,\theta^*)\cap T(\Theta_0,\theta^*)^{\perp}$. An exact expression of the latter can be derived in the general case of a block-diagonal covariance matrix $\Gamma$, while the former is not always available in a closed form, especially for nonlinear mixed effects models.

\subsubsection{Estimation of the Fisher Information Matrix}
For linear mixed effects models, the Fisher Information Matrix can be computed exactly. However it is not the case for nonlinear mixed effects models. The existing \proglang{R} packages provide in some cases an approximation of the FIM based for example on a linearisation of the model, or on a block-diagonal approximation of the FIM where a zero correlation is assumed between fixed effects and variance components. Nevertheless, the FIM is in general non block-diagonal. Moreover  these estimates are in general only valid for large samples, leading to possible wrong conclusions when considering small samples. Adapted tools have to be developed for such settings.

In the \pkg{varTestnlme} package, we provide three options for estimating the FIM, the first being specifically adapted for small samples sizes.  This approach is more time consuming but more accurate, in particular for small samples. It consists in building an estimate of the FIM via parametric bootstrap. Let us denote by $\hat{\theta}_1$ the estimate of $\theta$ obtained for the alternative model under $H_1$, with \pkg{lme4}, \pkg{nlme} or \pkg{saemix}. We generate $B$ bootstrap samples $(\varphi_i^{b,*}, y_i^{b,*}), 1\leq i \leq n, 1 \leq b \leq B$ from the alternative model, using the parameter value $\hat{\theta}_1$ and equations \eqref{eq:intravar}-\eqref{eq:inter_var}. Then, for each of these bootstrap samples, we compute the associated estimate $\hat{\theta}_b^*$. The bootstrap estimate of the FIM is then given by the empirical covariance matrix between the bootstrap estimates:
\begin{equation}
\hat{I}^* = \frac{1}{B} \sum_{b=1}^B \hat{\theta}_b^* (\hat{\theta}_b^*)^\top - \bar{\hat{\theta}}^* (\bar{\hat{\theta}}^*)^\top,
\end{equation}
where $\bar{\hat{\theta}}^* = \frac{1}{B} \sum_{b=1}^B \hat{\theta}_b^*$.

The  second  option consists in using  the FIM estimate as it is computed by the package that was used to fit the models, when possible.
Finally, the third option is to provide a user-defined positive semidefinite matrix which should be an estimate of the FIM. It can be for example computed by using the R package PFIM 4.0 (\cite{dumont2018pfim}) based on Monte Carlo tools or by implementing the method presented in (\cite{delattre2019estimating}) based on stochastic approximation tools.

\subsubsection{Computation of the tangent cone}
In \cite{bck19}, it was highlighted that the limiting distribution of the likelihood ratio test statistic depends on the structure of $\Gamma$ and  its expression is exhibited in two common cases, when $\Gamma$ is either diagonal or full. In the general case, $\Gamma$ can always be written in the following block-diagonal form:
\begin{equation}
\Gamma = \begin{pmatrix}
\Gamma_1 & 0 & \dots & 0 \\
0 & \Gamma_2 & \dots & 0 \\
\vdots & \ddots & \ddots & \vdots\\
0 & \dots & 0 & \Gamma_K
\end{pmatrix},
\end{equation}
where $K \in \mathbb{N}^*$, and for all $1 \leq k \leq K$,  $\Gamma_k$ is a positive semi-definite matrix of size $r_k \times r_k$ with $\sum_1^K r_k=p$. In the sequel, we assume that the blocks $\Gamma_1, \dots, \Gamma_K$ are full i.e., all the covariances inside these blocks are non-null. This is how the covariance matrix of random effects is implemented in package \pkg{lme4}  \cite[p. 7, section 2.2]{lme4}. It the appendix of this book, it is mentioned that more general structures for the covariance matrix $\Gamma$ could be implemented using modularized functions of the \pkg{lme4} package. Nevertheless these functionalities are not treated in the \pkg{varTestnlme} package at the moment. In \pkg{nlme}, more sophisticated covariance structures can be used, and in \pkg{saemix}, no specific restriction is imposed on the covariance matrix structure. However in \pkg{varTestnlme} we restrict ourselves to the block-diagonal case with full blocks. This restriction allows for a simple block-diagonal structure of the covariance matrix and eases the computation. 

In this context of a block-diagonal covariance matrix $\Gamma$, three cases can arise when testing that a subset of the variances are null: i) we are testing that $R$ blocks among the $K$ blocks $\Gamma_1, \dots, \Gamma_K$ are null, ii) we are testing that $R$ sub-blocks of the $K$ blocks $\Gamma_1, \dots, \Gamma_K$ are null, or iii) a mixture of i) and ii). 
Let us denote by $\ell_0$ the number of blocks into which covariances are tested equal to 0 without testing that the corresponding variances are equal to 0 (only testing non-diagonal elements), by $\ell_1$ the number of blocks which are tested entirely equal to 0, and by $\ell_2$ the number of blocks where sub-blocks are tested equal to 0. By sub-block, we mean that we are testing a sub-matrix which is strictly smaller than the block matrix from which it was extracted. 
We assume that $0 \leq \ell_0 \leq K$, $0 \leq \ell_1 \leq K$, $0 \leq \ell_2 \leq K$ and $1 \leq \ell_0 + \ell_1 + \ell_2 \leq K$.

Without loss of generality, and up to a permutation of rows and columns of $\Gamma$, we can assume that the blocks are in the following order: first, the blocks which are not tested, then the blocks where only covariances are tested, next the blocks which are tested entirely equal to zero, and finally the blocks where diagonal sub-blocks are tested equal to zero. Then, the null and alternative hypotheses can be written as: 
\begin{equation}
H_0 : \theta \in \Theta_0 \quad \text{against} \quad H_1 : \theta \in \Theta,
\end{equation}
where 
\begin{align}
\label{eq:hyp0} \Theta_0= \{ & \theta \in \mathbb{R}^q \mid \ \beta \in \mathbb{R}^b,  \\
\nonumber & \ \forall k = 1, \dots, K-(\ell_0+\ell_1+\ell_2), \Gamma_k \in \mathbb{S}^{r_k}_+, \\
\nonumber & \ \forall k = K-(\ell_0+\ell_1+\ell_2)+1, \dots, K-(\ell_1+\ell_2), \Gamma_k \in \mathbb{S}^{r_k}_+, \text{ with } t_k \text{ covariances null}, \\
\nonumber & \ \forall k = K-(\ell_1+\ell_2)+1, \dots, K-\ell_2, \Gamma_k = 0,\\
\nonumber & \ \forall k = K-\ell_2+1, \dots, K, \Gamma_k  = \left( \begin{array}{c|c}
\tilde{\Gamma}_k &  0 \\
\hline
0 & 0
\end{array}
\right), \text{ with } \tilde{\Gamma}_k \in \mathbb{S}^{r_k-s_k}_{+},\\
\nonumber & \ \Sigma \in \mathbb{S}^J_+
\}\\[0.3cm]
\label{eq:hyp1} \Theta = \{ & \theta \in \mathbb{R}^q \mid  \ \beta \in \mathbb{R}^b,  \forall k = 1, \dots, K, \Gamma_k \in \mathbb{S}^{r_k}_+, \Sigma \in \mathbb{S}^J_+\}
\end{align}

From the general expressions of $\Theta_0$ and $\Theta$ in \eqref{eq:hyp}, we can derive the expressions of $T(\Theta,\theta^*)$ and $T(\Theta_0,\theta^*)$ using the results of \cite{Hir12}, and consequently the expression of the closed convex cone $\mathcal{C}:= T(\Theta,\theta^*)\cap T(\Theta_0,\theta^*)^{\perp}$, involved in the chi-bar-square distribution in \eqref{eq:lrtdist}.

We obtain:
\begin{align}\label{eq:cone}
\mathcal{C} = & \{0\}^{a} \times \underbrace{\mathbb{R}^{r_f}}_{\substack{\textrm{tested} \\ \textrm{fixed effects}}} \times \underbrace{\prod_{k=K-(\ell_0+\ell_1+\ell_2)+1}^{K-(\ell_1+\ell_2)} \mathbb{R}^{t_k}}_{\substack{\textrm{covariances tested without testing} \\[0.1cm] \textrm{the corresponding variances}}} \times \underbrace{\prod_{k=K-(\ell_1+\ell_2)+1}^{K-\ell_2} \mathbb{S}^{r_k}_+}_\textrm{fully tested blocks} \times\underbrace{\prod_{k=K-\ell_2+1}^K \{\mathbb{R}^{s_k(r_k-s_k)} \times \mathbb{S}^{s_k}_+ \}}_{\substack{\textrm{sub-blocks tested}}},
\end{align}
where
\begin{align}\label{eq:dim0}
a & = \underbrace{b-r_f}_{\substack{\textrm{non tested} \\ \textrm{fixed effects}}} + \underbrace{\sum_{k=1}^{K-(\ell_0+ \ell_1+\ell_2)} \frac{r_k(r_k-1)}{2}}_\textrm{non tested blocks} +\underbrace{\sum_{k=K-\ell_2+1}^K \frac{(r_k-s_k)(r_k-s_k+1)}{2}}_{\substack{\textrm{untested sub-blocks in blocks} \\[0.1cm] \textrm{where a sub-block is tested}}} + \underbrace{\frac{J(J+1)}{2}}_{\substack{\textrm{residual} \\[0.1cm] \textrm{covariance matrix}}}.
\end{align}

To identify the components of the chi-square mixture i.e., the degrees of freedom and the weights of all the chi-square distributions involved in the mixture, we can use the properties enounced by \cite{Sha85}, stating that if $\mathcal{C}$ contains a linear space of dimension $d_1$, the first $d_1$ weights of the mixture are null, and if $\mathcal{C}$ is included in a linear space of dimension $d_2$, the last $(q-d_2)$ weights of the mixture are null. The chi-bar-square distribution $\bar{\chi}^2(I^{-1}_*,\mathcal{C})$ is then a mixture of $(d_2-d_1+1)$ chi-square distributions with degrees of freedom varying between $d_1$ and $d_2$.

According to the general formulation of the cone $\mathcal{C}$ given in \eqref{eq:cone}, we have:
\begin{gather}
\begin{aligned}\label{eq:dimsubspaces}
	d_1 & = r_f + \sum_{k=K-(\ell_0+\ell_1+\ell_2)+1}^{K-(\ell_1+\ell_2)} t_k +  \sum_{k=K-\ell_2+1}^K s_k(r_k-s_k), \\
	d_2 & = q-a.
\end{aligned}
\end{gather}

In particular, if only blocks of variances are tested (i.e., if $r_f = 0$, $\ell_0 = 0$ and $\ell_2=0$), then $d_1=0$ and there is a Dirac mass at 0 in the mixture.

From equations \eqref{eq:dim0} and \eqref{eq:dimsubspaces}, we can see that the number of elements in the chi-bar-square mixture only depends on the number of variances being tested, and on the structure of the covariance matrix. When $r_f$ fixed effects are tested simultaneously to a set of variance components, the number of elements in the mixture is the same as in the case where only the set of variance components is tested, but the degrees of freedom of each element of the chi-square mixture is shifted upward by $r_f$.

As an example, let us consider a model with 3 random effects, with $\varphi_i = \beta + b_i$, $b_i \sim \mathcal{N}(0, \Gamma)$, $\varepsilon_i \sim \mathcal{N}(0, \Sigma)$ with $\Gamma$ and $\Sigma$ positive definite covariance matrices. Let us consider the following hypotheses: 
\begin{equation}
H_0 : \theta \in \Theta_0 \quad \text{against} \quad H_1 : \theta \in \Theta,
\end{equation}
where 
\[
\Theta_0=\{\beta \in \mathbb{R}^3, \Gamma = \left( \begin{array}{ccc}
\gamma_1^2 &  0 & 0 \\
0 & 0 & 0 \\
0 & 0 & 0
\end{array}\right), \Sigma \in \mathbb{S}^3_+\} \text{ and } 
\Theta=\{\beta \in \mathbb{R}^3, \Gamma = \left( \begin{array}{ccc}
\gamma_1^2 &  \gamma_{12} & \gamma_{13} \\
\gamma_{12} & \gamma_2^2 & \gamma_{23} \\
\gamma_{13} & \gamma_{23} & \gamma_3^2
\end{array}\right), \Sigma \in \mathbb{S}^3_+ \}
\]

Since no fixed effect is tested, we have $r_f = 0$. We can consider $\Gamma$ as a block-diagonal matrix with only $K=1$ block. In this block of size $3\times 3$, we are testing that a sub-block of size $2\times 2$ is equal to 0. Using the above notations, we have $\ell_0=0, \ell_1=0$ and $\ell_2=1$, $r_1 = 3$ and $s_1=2$, and $J=3$. We thus have: $a = 3 - 0 + 0 +\frac{(3-2)(3-2+1)}{2} + 6 = 10$, $d_1 = 0 + 1(3-1) = 2$ and $d_2 = 15-10=5$. The limiting distribution of the LRT associated to the two above hypotheses is then a mixture of $d_2-d_1+1=4$ chi-square distributions with degrees of freedom 2, 3, 4 and 5.


\subsection{Computation of the chi-bar-square weights}\label{sec:weight}
The chi-bar-square weights are in general non explicit, and have to be estimated via Monte Carlo techniques. The general idea is to simulate $M$ i.i.d. realizations $X_1, \dots, X_M$ from the limiting chi-bar-\-square distribution using its definition as the norm of the projection of a multivariate Gaussian random variable on a closed convex cone.
More precisely, let $\mathcal{C}$ be a closed convex cone of $\mathbb{R}^q$, $V$ a positive definite matrix of size $q \times q$ and $Z \sim \mathcal{N}(0,V)$. Then, the random variable $X$ defined below follows a chi-bar-square distribution with parameters $V$ and $\mathcal{C}$ as detailed in \cite{Sil11}:
\begin{equation}\label{eq:chibarsquareRV}
X = Z^\top V^{-1}Z-\min_{\theta \in \mathcal{C}}(Z-\theta)^\top V^{-1}(Z-\theta).
\end{equation}

In \pkg{varTestnlme}, when the weights of the chi-bar-square distribution are not available in a closed form, they are estimated according to the procedure proposed by \cite{Sil11}, using an estimate of the Fisher Information Matrix. More details are given in Algorithm \ref{alg:simuCBS}.

\begin{algorithm}
  \caption{Estimation of the weights of the chi-bar-square distribution using Monte Carlo simulations\label{alg:simuCBS}}
  \begin{algorithmic}[1]
  \State \textbf{Define} $\hat{I}^{-1}$ an estimate of the inverse of the Fisher information matrix  under $H_1$, and let $c_{d_1+2} <  \dots < c_{q-d_2}$ be a sequence of non-negative increasing numbers.
  \For {$i = 1, \dots, M$}
  			\begin{enumerate}
  				\item simulate $Z_i \sim \mathcal{N}(0,\hat{I}^{-1})$ 
  				\item compute $$X_i := Z_i^\top \hat{I}^{-1}Z_i-\min_{\theta \in \mathcal{C}}(Z_i-\theta)^\top \hat{I}^{-1}(Z_i-\theta),$$ using quadratic programming when $\mathcal{C}$ is the non-negative orthant $\mathbb{R}^r$, for some $r \leq q$, and using general nonlinear optimization tools otherwise. In \pkg{varTestnlme}, we use respectively the \pkg{quadprog} and the \pkg{alabama} packages.
  			\end{enumerate}
  \EndFor
  \State Compute matrix $A$ of size $(q-d_2) \times (q-d_2)$:
	\begin{equation*}
		A = \begin{pmatrix}
					1 & 1 & 1 & \dots & 1 & 1 \\
					1 & 0 & 1 & \dots & 1 & 0 \\
					F_{d_1}(c_{d_1+2}) & F_{d_1+1}(c_{d_1+2}) & F_{d_1+2}(c_{d_1+2}) & \dots & F_{q-d_2-1}(c_{d_1+2}) & F_{q-d_2}(c_{d_1+2}) \\
					\vdots & \vdots & \vdots & \cdots & \vdots & \vdots \\
					F_{d_1}(c_{q-d_2}) & F_{d_1+1}(c_{q-d_2}) & F_{d_1+2}(c_{q-d_2}) & \dots & F_{q-d_2-1}(c_{q-d_2}) & F_{q-d_2}(c_{q-d_2}) 	
				\end{pmatrix},
	\end{equation*}    
	 where $F_i$ is the cumulative distribution function of a chi-square distribution with $i$ degrees of freedom.
	 
\noindent	 \textit{N.B. The expression of $A$ given above corresponds to the case where $(q-d_2)$ is even.  In the case where $(q-d_2)$ is odd, the last column has a `1' in the second row, instead of a `0'.}

  \State Compute vector $\hat{b}$ of size $(q-d_2)$ :
	\begin{equation*}
		\hat{b} = \begin{pmatrix}
		1 \\
		1/2 \\
		\frac{1}{M} \sum_{i=1}^M \mathbb{1}_{X_i \leq c_{d_1+2}}\\
		\vdots \\
		\frac{1}{M} \sum_{i=1}^M \mathbb{1}_{X_i \leq c_{q-d_2}}
		\end{pmatrix}
	\end{equation*}	  
  \State Estimate the weights of the chi-bar-square distribution by solving the system:
  $$A\hat{w} = \hat{b} \Leftrightarrow \hat{w} = A^{-1} \hat{b}$$
  \State Estimate the covariance matrix of the weights by:
  $$\text{Var}(\hat{w}) = A^{-1} \ \text{Var}(\hat{b}) \ (A^{-1})^\top$$
  \end{algorithmic}
\end{algorithm}

\subsection{Computation of the $p$-value}
Let $F_d$ be the cumulative distribution function (cdf) of the chi-square distribution wih $d$ degrees of freedom. The $p$-value of the test can then be estimated in two different ways, both of them being computed in the \pkg{varTestnlme}:
\begin{align}
\label{eq:pvalw} \hat{p}_1 & = \sum_{j=1}^{q+1-d_1-d_2} \hat{w}_j \ (1-F_{d_1+j-1}(LRT_n)), \\
\label{eq:pvalsamp} \hat{p}_2 & = \frac{1}{M} \sum_{i=1}^M \mathbb{1}_{X_i \geq LRT_n},
\end{align}
where $\hat{w}_j$ is the estimated weight associated with the chi-square distribution with $(d_1+j-1)$ degrees of freedom, and where $X_1,\dots,X_M$ are simulated according to the limiting chi-bar-square distribution (see Algorithm \ref{alg:simuCBS} for more details on the notations).

Note that for very small values of the real $p$-value, a very large sample size $M$ would be needed in order to get a non-zero estimate $\hat{p}_2$. 

\subsection{Bounds on $p$-values}\label{sec:boundpval}
Since the simulation of $X_1, \dots, X_M$ can be time consuming, and since $F_i(c) \geq F_j(c)$ for $i < j$, it is possible to compute bounds on the $p$-value of the test. Indeed, since the sum of all the weights is equal to 1, and the sum of even (resp. odd) weights is equal to 1/2, natural bounds are given for the $p$-value by \cite{Sil11}:
\begin{equation}\label{eq:bounds}
	1- \frac{F_{d_1}(LRT_n) + F_{d_1+1}(LRT_n)}{2} \leq p \leq 	1- \frac{F_{q-d_2-1}(LRT_n) + F_{q-d_2}(LRT_n)}{2}.
\end{equation}

Even if those bounds can be crude in some cases, they can turn out to be quite useful in practice, and might bring enough information to support or reject the null hypothesis.

By default, package \pkg{varTestnlme} returns these bounds, and if more precision is wanted by the user, it is possible to re-run the analysis in order to estimate the weights of the limiting distribution.



\section{Illustrations} \label{sec:illustrations}
The \pkg{varTestnlme} package provides a unified framework for likelihood ratio tests of fixed and random parts of a linear, generalized linear or nonlinear mixed-effects model fitted either with the \pkg{nlme}, \pkg{lme4} or \pkg{saemix} packages. The main function \code{varCompTest} takes, in its simplest form, two arguments: \code{m1} the fitted model under the alternative $H_1$ and \code{m0} the fitted model under the null $H_0$. 

\subsection{Data}
For illustrative purposes, three datasets will be used in the paper, each of them covering one of the three types of models that can be treated with the \pkg{varTestnlme} package. These datasets are all available in the \pkg{nlme} package. 

\subsubsection{Orthodontal data}
\begin{figure}
\centering
\includegraphics{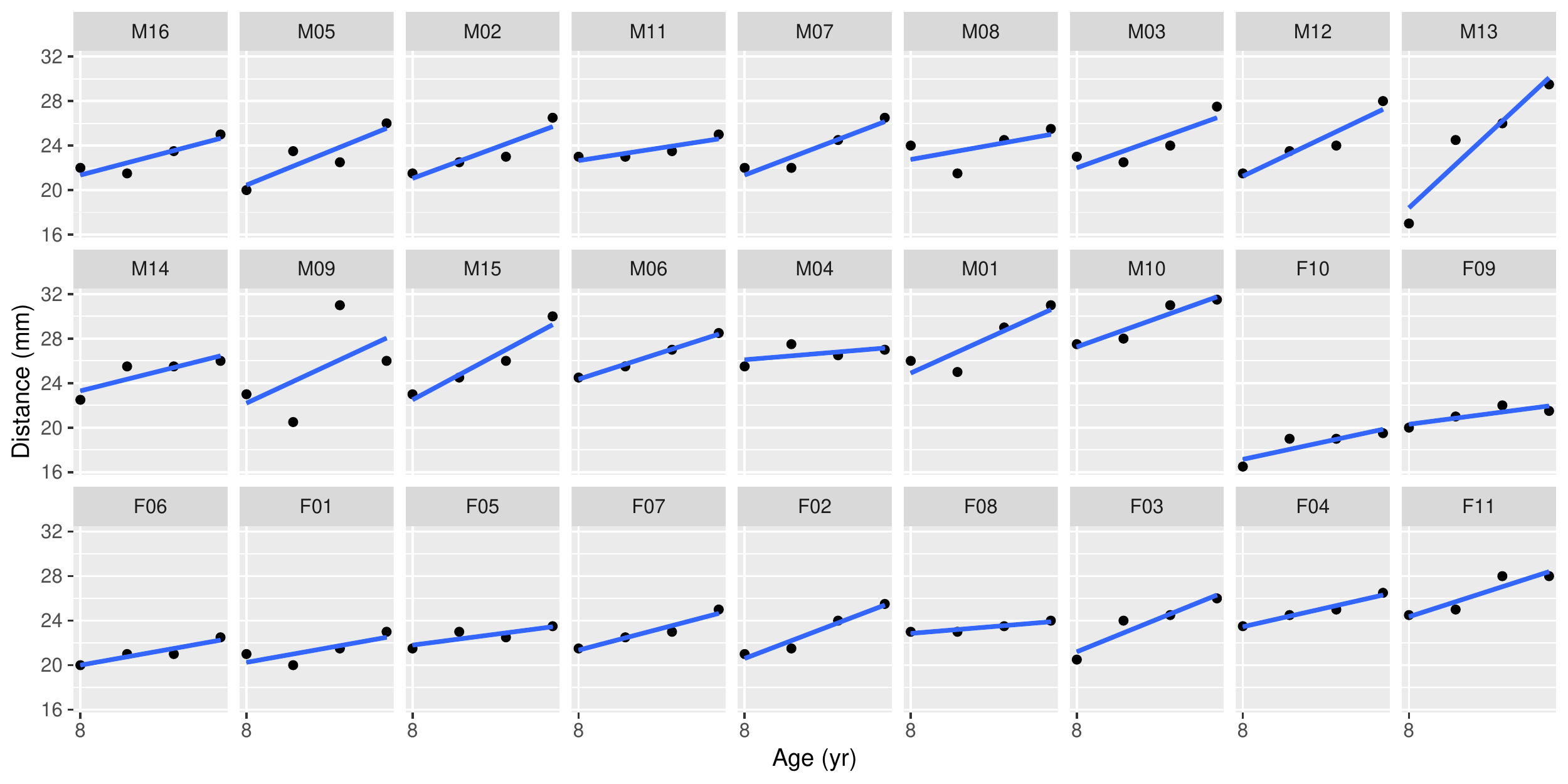}
\caption{Data from the orthodontal study: distance between the pituitary and the pterygomaxillary fissure as a function of age, by subject.}
\label{fig:dental}
\end{figure}

The first dataset comes from a study on dental growth \citep{Pott64}, where the distance between the pituitary and the pterygomaxillary fissure was recorded every two years from the age of 8 to the age of 14, on 27 children, 16 boys and 11 girls (see Figure \ref{fig:dental}). 

\begin{CodeChunk}
\begin{CodeInput}
R>  data("Orthodont", package = "nlme")
\end{CodeInput}
\end{CodeChunk}

These data can be fitted using a linear mixed-effects model, with a random slope and a random intercept. 
Let us denote by $y_{ij}$, $1 \leq i \leq 27$, $1 \leq j \leq 4$ the dental measurement of child $i$ of sex $x_i$ at age $t_j$. Then the model can be written as:
\begin{align}\label{eq:linmod}
y_{ij} & = (\beta_1 + \beta_2 x_i + b_{i1}) + (\beta_3 + \beta_4 x_i + b_{i2}) t_j + \varepsilon_{ij} \  , \\
\varepsilon_{ij} & \sim \mathcal{N}(0,\sigma^2), \quad (b_{i1},b_{i2})^\top \sim \mathcal{N}(0,\Gamma) \ .
\end{align}
We denote by $\beta = (\beta_1, \beta_2, \beta_3, \beta_4)^\top$ the vector of fixed effects.

\subsubsection{Bovine pleuropneumonia}
\begin{figure}
\centering
\includegraphics{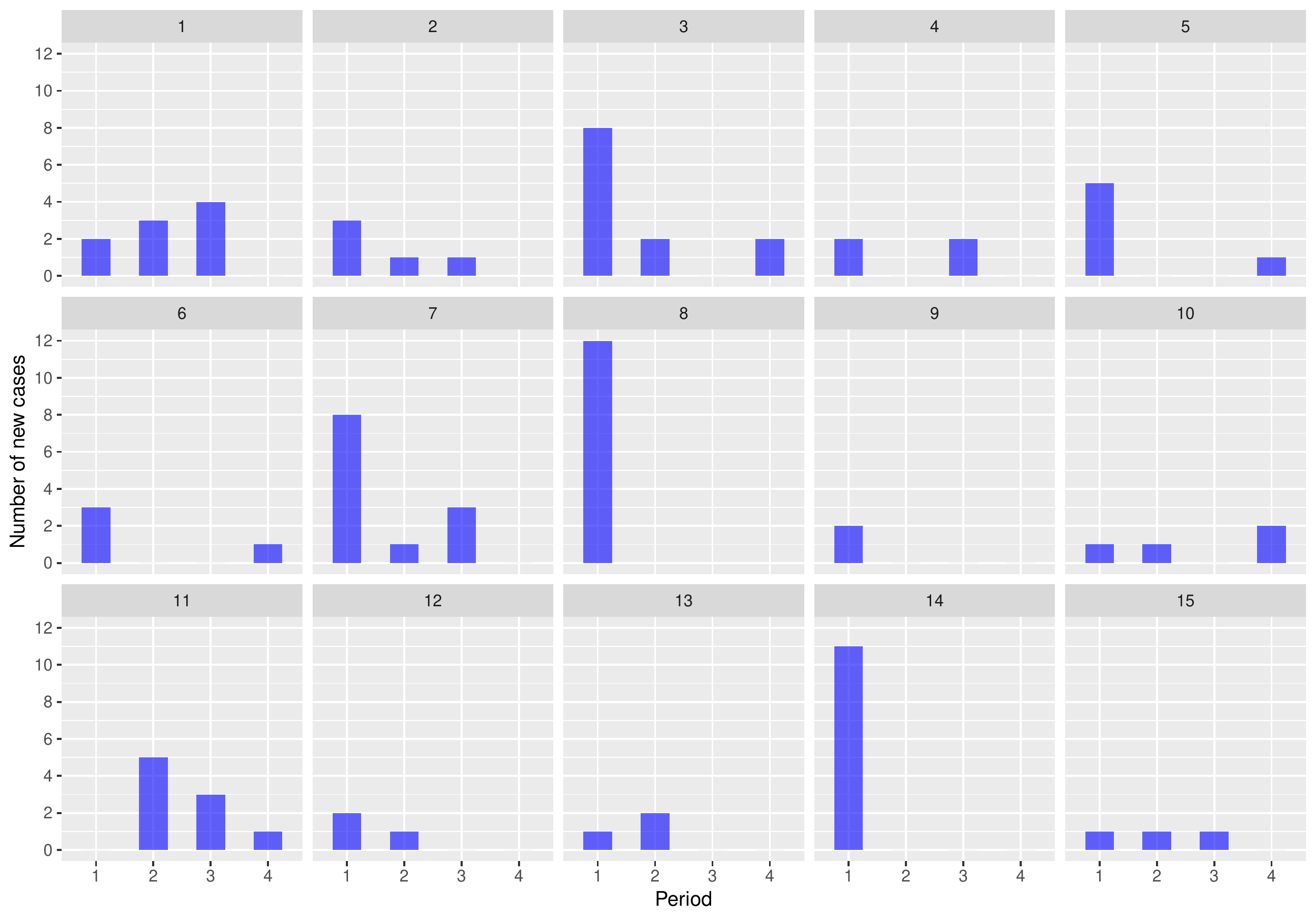}
\caption{Data from the cbpp study: number of new cases per period, by herd.}
\label{fig:cbpp}
\end{figure}

The second dataset comes from a study on contagious bovine pleuropneumonia (cbpp) \citep{Les04}, where the number of new serological cases occuring during a given time period was recorded at 4 occasions, on 15 herds (see Figure \ref{fig:cbpp}). 

\begin{CodeChunk}
\begin{CodeInput}
R>  data("cbpp", package = "nlme")
\end{CodeInput}
\end{CodeChunk}

This data can be fitted using a generalized linear mixed-effects model and a logistic regression. Let us denote by $y_{ij}$ the number of serological cases in herd $i$ and at time  period $t_j$. If we denote by $g$ the logit function, then the model is given by:
\begin{align}\label{eq:glinmod}
\mathbb{E}(y_{ij}) & = g^{-1}(b_{0,i} + \beta_1  t_j),\\
b_{0,i} & \sim \mathcal{N}(0,\gamma^2)
\end{align}

\subsubsection{Loblolly pine trees}

\begin{figure}
\centering
\includegraphics{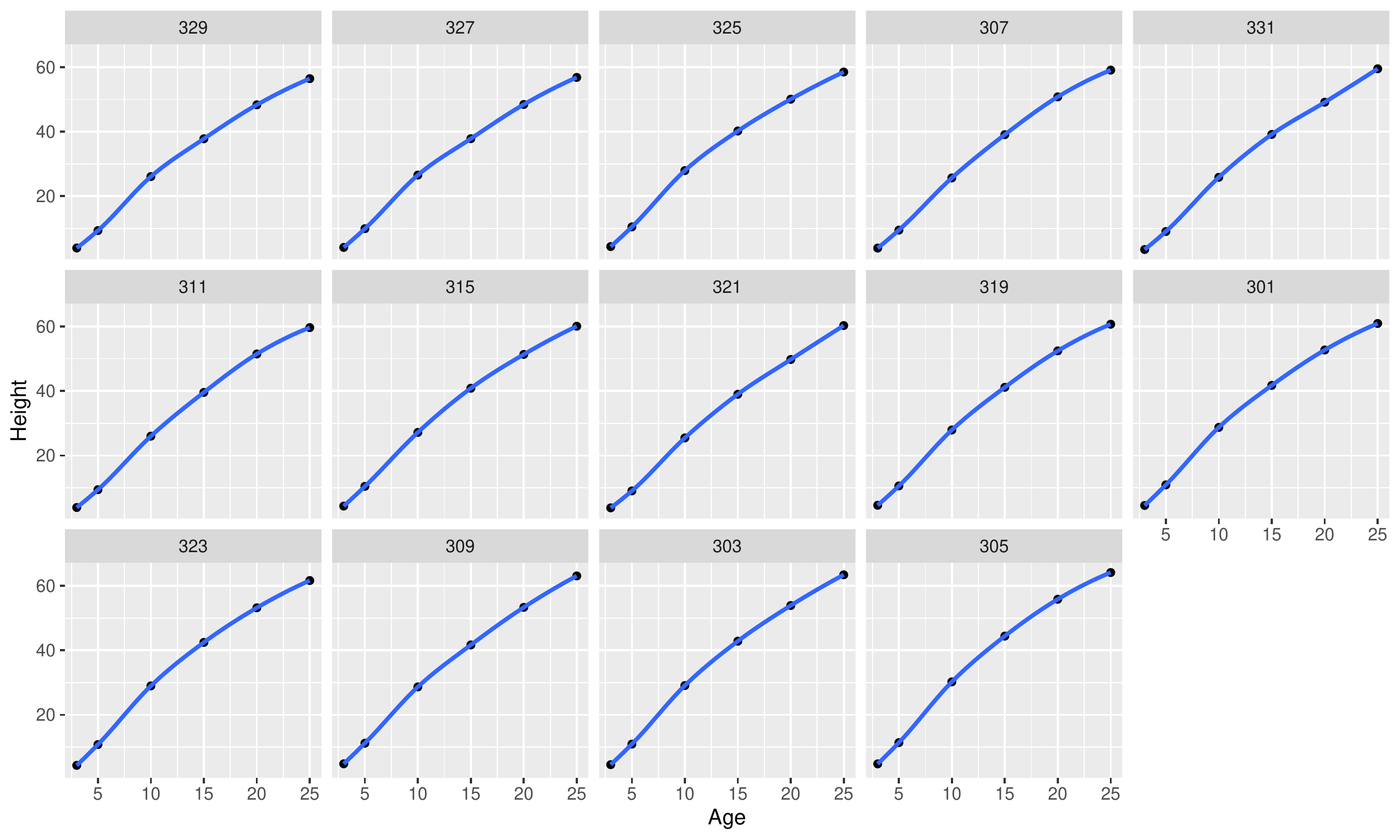}
\caption{Data from the loblolly study: height  as a function of age, by tree.}
\label{fig:loblolly}
\end{figure}

The third dataset comes from a study on Loblolly pine trees \citep{Kung86}, where the growth of 14 trees was recorded between 3 and 25 years of age (see Figure \ref{fig:loblolly}).

\begin{CodeChunk}
\begin{CodeInput}
R>  data("Loblolly", package = "datasets")
\end{CodeInput}
\end{CodeChunk}

This data can be fitted using a nonlinear mixed-effects model. Let us denote by $y_{ij}, 1 \leq i \leq 14, 1 \leq j \leq 6$ the height of the $i$-th tree at age $x_j$, by $Asym_i$ the asymptote for the $i$-th tree, $lrc_i$ the logarithm of the growth rate for the $i$-th tree and $R_{0,i}$ the height of tree $i$ at age 0. We consider the following model:
\begin{align}\label{eq:nonlinmod}
	y_{ij} & = Asym_{i} + (R_{0,i} - Asym_{i}) \exp(- e^{lrc_{i}} x_j ) + \varepsilon_{ij}, 	\ \varepsilon_{ij} \sim \mathcal{N}(0,\sigma^2)\\ 
	(Asym_i,R_{0,i},lrc_i)^\top & = \beta + b_i, \ b_i \sim \mathcal{N}_3(0, \Gamma)
\end{align}

\subsection{Preliminary step: fitting the models under $H_0$ and $H_1$}\label{sec:illustr_prel}
The first step of the analysis consists in specifying $H_0$ and $H_1$, the two hypotheses defining the test. Then, one needs to fit the model under both  hypotheses, using one of the following three packages: \pkg{nlme}, \pkg{lme4} and \pkg{saemix}. The \pkg{varTestnlme} package automatically detects  
the structure of the models under $H_0$ and $H_1$. 

\emph{Note that both models \code{m1} and \code{m0} should be fitted with the same package, except when there is no random effects in \code{m0}} (see details below).

\subsubsection{Linear model}
We will consider three hypothesis testing configurations for the linear model presented in \eqref{eq:linmod}. We detail below the null and alternative hypotheses in each case, and the code to fit the associated models.
\begin{enumerate}
	\item \textbf{\textsc{Case 1:}} testing that there is a random slope, in a model where the slope and the intercept are correlated:
	$$H_0: \theta \in \Theta_0 \quad \text{against} \quad H_1: \theta \in \Theta,$$
	with
	$$\Theta_0 = \{\beta \in \mathbb{R}^4, \Gamma = \begin{pmatrix}
	\gamma_1^2 & 0 \\
	0 & 0
	\end{pmatrix}, \sigma^2 > 0 \} \text{ and } \Theta = \{\beta \in \mathbb{R}^4, \Gamma = \begin{pmatrix}
	\gamma_1^2 & \gamma_{12} \\
	\gamma_{12} & \gamma_2^2
	\end{pmatrix}, \sigma^2 > 0\}.$$
The syntax using \pkg{lme4} and \pkg{nlme} packages is:
		\suspend{enumerate}

\begin{CodeChunk}
\begin{CodeInput}
R>  lm1.h1.lme4 <- lmer(distance ~ 1 + Sex + age + age * Sex + 
+    (1 + age | Subject), data = Orthodont, REML = FALSE)
R>  lm1.h0.lme4 <- lmer(distance ~ 1 + Sex + age + age * Sex +  
+   (1 | Subject),  data = Orthodont, REML = FALSE)
\end{CodeInput}
\end{CodeChunk}

\begin{CodeChunk}
\begin{CodeInput}
R>  lm1.h1.nlme <- lme(distance ~ 1 + Sex + age + age * Sex, 
+    random = ~ 1 + age | Subject, data = Orthodont, method = "ML")
R>  lm1.h0.nlme <- lme(distance ~ 1 + Sex + age + age * Sex, 
+    random = ~ 1 | Subject, data = Orthodont, method = "ML")
\end{CodeInput}
\end{CodeChunk}
	
\resume{enumerate}
		\item \textbf{\textsc{Case 2:}} testing that there is a random slope, in a model where  the slope and the intercept are independent:		  
			$$H_0: \theta \in \Theta_0 \quad \text{against} \quad H_1: \theta \in \Theta,$$
	with
		$$\Theta_0 = \{\beta \in \mathbb{R}^4, \Gamma = \begin{pmatrix}
	\gamma_1^2 & 0 \\
	0 & 0
	\end{pmatrix}, \sigma^2 > 0 \} \text{ and } \Theta = \{\beta \in \mathbb{R}^4, \Gamma = \begin{pmatrix}
	\gamma_1^2 & 0 \\
	0 & \gamma_2^2
	\end{pmatrix}, \sigma^2 > 0\}$$
	\suspend{enumerate}

\begin{CodeChunk}
\begin{CodeInput}
R>  lm2.h1.lme4 <- lmer(distance ~ 1 + Sex + age + age * Sex + 
+   (1 + age || Subject),  data = Orthodont, REML = FALSE)
R>  lm2.h0.lme4 <- lm1.h0.lme4

R>  lm2.h1.nlme <- lme(distance ~ 1 + Sex + age + age * Sex, 
+    random =  list(Subject = pdDiag(~1+age)), data = Orthodont, method = "ML")
R>  lm2.h0.nlme <- lm1.h0.nlme
\end{CodeInput}
\end{CodeChunk}

	\resume{enumerate}	
		\item \textbf{\textsc{Case 3:}} testing for the presence of random effects, in a model where the slope and the intercept are independent:	
		$$H_0: \theta \in \Theta_0 \quad \text{against} \quad H_1: \theta \in \Theta,$$
	with
		$$\Theta_0 = \{\beta \in \mathbb{R}^4, \Gamma = \begin{pmatrix}
	0 & 0 \\
	0 & 0
	\end{pmatrix}, \sigma^2 > 0 \} \text{ and } \Theta= \{\beta \in \mathbb{R}^4, \Gamma = \begin{pmatrix}
	\gamma_1^2 & 0 \\
	0 & \gamma_2^2
	\end{pmatrix}, \sigma^2 > 0\}$$
\end{enumerate}

\begin{CodeChunk}
\begin{CodeInput}
R>  lm3.h1.lme4 <- lm2.h1.lme4
R>  lm3.h1.nlme <- lm2.h1.nlme
R>  lm3.h0 <- lm(distance ~ 1 + Sex + age + age * Sex, data = Orthodont)
\end{CodeInput}
\end{CodeChunk}	

\subsubsection{Generalized linear model}
In the model considered in \eqref{eq:glinmod}, there is only one random effect.  We will then consider the following test:
	$$H_0: \theta \in \Theta_0 \quad \text{against} \quad H_1: \theta \in \Theta,$$
	with
\[
\Theta_ 0 = \{\beta \in \mathbb{R}, \gamma = 0 \} \text{ and } \Theta= \{\beta \in \mathbb{R}, \gamma \geq 0\}.
\]

The corresponding code using \pkg{lme4} is:
\begin{CodeChunk}
\begin{CodeInput}
R>  glm1 <- glmer(cbind(incidence, size - incidence) ~ period + (1 | herd), 
	     family = binomial, data = cbpp)
R>  glm0 <- glm(cbind(incidence, size - incidence) ~ period,
             family = binomial, data = cbpp)
\end{CodeInput}
\end{CodeChunk}	

\subsubsection{Nonlinear model}
Let us consider the nonlinear model defined in \eqref{eq:nonlinmod}.
Here we will carry out the following test:
	$$H_0: \theta \in \Theta_0 \quad \text{against} \quad H_1: \theta \in \Theta,$$
	with
\[
\Theta_0 = \{\beta \in \mathbb{R}^3, \Gamma = \left( \begin{array}{ccc}
\gamma_1^2 &  0 & 0 \\
0 & 0 & 0 \\
0 & 0 & 0
\end{array}\right), \sigma > 0\} \text{ and } 
\Theta = \{\beta \in \mathbb{R}^3, \Gamma = \left( \begin{array}{ccc}
\gamma_1^2 &  0 & 0 \\
0 & \gamma_2^2 & 0 \\  
0 & 0 & \gamma_3^2
\end{array}\right), \sigma > 0 \},
\]
i.e., that only the asymptote is random.

The corresponding code using \pkg{nlme} and \pkg{lme4} is:
\begin{CodeChunk}
\begin{CodeInput}
R>  start <- c(Asym = 103, R0 = -8.5, lrc = -3.2)
R>  nlm1.nlme <- nlme(height ~ SSasymp(age, Asym, R0, lrc),
                 fixed = Asym + R0 + lrc ~ 1,
                 random = pdDiag(Asym + R0 + lrc ~ 1),
                 start = start,
                 data = Loblolly)
R>  nlm0.nlme <- nlme(height ~ SSasymp(age, Asym, R0, lrc),
                 fixed = Asym + R0 + lrc ~ 1,
                 random = pdDiag(Asym ~ 1),
                 start = start,
                 data = Loblolly)
R>  nlm1.lme4 <- nlmer(height ~ SSasymp(age, Asym, R0, lrc) 
		 ~ (Asym + R0 + lrc || Seed), 
		 start = start,
		 data = Loblolly)
R>  nlm0.lme4 <- nlmer(height ~ SSasymp(age, Asym, R0, lrc) 
		 ~ (Asym | Seed), 
		 start = start,
		 data = Loblolly)
\end{CodeInput}
\end{CodeChunk}	                 

Using \pkg{saemix}, the syntax is given by:
\begin{CodeChunk}
\begin{CodeInput}
R>  modelSSasymp <- function(psi, id, xidep){
	Asym <- psi[id,1]
	R0 <- psi[id,2]
	lrc <- psi[id,3]
	age <- xidep
	ypred <- Asym + (R0 - Asym) * exp( - exp(lrc) * age)
	return(ypred)}
R>  nlm1.saemix <- saemixModel(model = modelSSasymp,
                               description = "Asymptotic regression",
                               psi0 = matrix(start, ncol = 3, byrow = TRUE,
                                      dimnames = list(NULL, c("Asym", "R0", "lrc"))),
                               transform.par = c(0,0,0),
                               fixed.estim = c(1,1,1),
                              covariance.model = matrix(c(1,0,0,
                                                          0,1,0,
                                                          0,0,1), ncol = 3),
                              omega.init = matrix(c(200,0,0,
                                                    0,750,0,
                                                    0,0,150), ncol = 3),
                              error.model = "constant")
R>  saemix.data <- saemixData(name.data = Loblolly, name.group = c("Seed"),
                              name.predictors = c("age"), name.response = c("height"),
                              units = list(x = "year",y = "feet"))
R>  saemix(saemix.model, saemix.data)
\end{CodeInput}
\end{CodeChunk}

\subsection{Variance component testing}
Once the models have been fitted using one of the three packages \pkg{lme4}, \pkg{nlme} or \pkg{saemix}, the test of the variance components can be run using the \fct{varCompTest} function of the \pkg{varTestnlme} package.
\begin{Code}
varCompTest(m1, m0, control = list(M=5000, parallel=F, nbcores=1, B=1000), 
             fim = "extract", pval.comp = "bounds")
\end{Code}
where:
\begin{itemize}
	\item \code{m1} is the model fitted under $H_1$ i.e., an object of class \code{merMod}, \code{glmerMod}, \code{lme}, \code{nlme} or \code{SaemixObject},
	
	\item \code{m0} is the model fitted under $H_0$ using the same package as for \code{m1}, except when no random effects are present under $H_0$ (when testing for the absence of random effects). In this case, \code{m0} should be fitted using \fct{lm} for linear, \fct{glm} for generalized linear, and \fct{nls} for nonlinear models (these functions are available in the \pkg{stats} package),
	
	\item \code{control} is a list with: \code{M}, the size of the Monte Carlo sample for the computation of the weights of the chi-bar-square distribution (\code{5000} by default, see Algorithm \ref{alg:simuCBS}); \code{parallel} a boolean to specify whether the Monte Carlo sampling should be done in parallel (\code{FALSE} by default); \code{nbcores} the number of cores to be used with parallel computing (\code{1} by default); and \code{B} the size of the bootstrap sample used to estimate the Fisher Information Matrix (\code{1000} by default),
	
	\item \code{fim} could be either \code{"extract"} (the default) if the Fisher Information Matrix should be extracted from the fitted model \code{m1}; \code{"compute"} if it should be computed using parametric bootstrap ; or \code{FIM} a user-defined matrix to be used as the Fisher Information Matrix,
	
	\item \code{pval.comp} specifies the way to compute the $p$-value, and could be either \code{"bounds"} (the default), in which case only bounds on the true $p$-value are computed, \code{"approx"}, in which case a Monte Carlo estimation of the exact $p$-value is provided, or \code{"both"} for a combination of both approaches. In the case where the weights are known explicitly, no approximation is made and the exact weights are return.
\end{itemize}

Note that the chi-bar-square weights approximation can be time consuming, especially when \code{control\$M} is high, or when the number of components in the chi-bar-square distribution increases. It is recommended to first run the function using the default setting, i.e., with \code{pval.comp="bounds"}, and to possibly re-run it with \code{pval.comp="approx"} if more precision is needed.

The function returns an object of classes \code{htest} and \texttt{vctest} and as such, can be printed using the provided \fct{print} function of \pkg{varTestnlme} or the \fct{print.htest} function of package \pkg{EnvStats}. In particular, it contains the following slots:
\begin{itemize}
	\item \code{null.value}: the value of the tested parameters under the null hypothesis,
	\item \code{alternative}: the value of the parameters under the alternative hypothesis,
	\item \code{statistic}: the value of the LRT statistic,
	\item \code{method}: the name of the statistical test,
	\item \code{parameters}: a list with the chi-bar-square distribution parameters: \code{df} the degrees of freedom, \code{weights} the weights associated with each component of the limiting chi-bar-square distribution, \code{sdWeights} the standard deviation associated with the estimation of each weights and \code{FIM} the estimate of the Fisher Information Matrix,
	\item \code{pvalue}: a named numeric vector with four elements: \code{pvalue.weights}, the $p$-value obtained using equation \eqref{eq:pvalw} (equals \code{NA} if weights were not computed, e.g., if option \code{pval.comp} was set to \code{"bounds"}), \code{pvalue.sample}: the $p$-value obtained using equation \eqref{eq:pvalsamp} (equals \code{NA} if weights were not computed or if exact weights were available and no sampling was done), \code{pvalue.lowerbound} and \code{pvalue.upperbound} the bounds on the $p$-value, obtained from equation \eqref{eq:bounds}.
\end{itemize}

\subsubsection{Linear model}
We illustrate the function on the three different tests defined in the previous section for the linear model on the orthodontal data.

\textsc{Case 1:} testing that the variance of age is null, in a model with correlated random effects. We first run the function with the default arguments. 

\begin{CodeInput}
R>  vt <- varCompTest(lm1.h1.lme4, lm1.h0.lme4)
R>  print(vt)
\end{CodeInput}
\begin{CodeOutput}
Variance components testing in mixed effects models
Testing that variance of age is null

 Likelihood ratio test statistic:
	LRT =  0.8326426

 p-value from estimated weights: 0.5104889
\end{CodeOutput}

Using the \fct{summary} function, we get the following output:
\begin{CodeInput}
R>  summary(vt)
\end{CodeInput}
\begin{CodeOutput}
Variance components testing in mixed effects models
Testing that variance of  age is null

 Likelihood ratio test statistic:
	LRT =  0.8326426

 Limiting distribution:
	mixture of 2 chi-bar-square distributions with degrees of freedom 1 2
	associated weights (and sd): 0.5 (0) 0.5 (0)

 p-value of the test:
	from estimated weights: 0.5104889
	bounds on p-value: lower  0.5104889 upper  0.5104889
\end{CodeOutput}

The LRT statistic is computed, and the asymptotic distribution is identified as a mixture between two chi-square distributions with degrees of freedom 1 and 2. In this case we can compute the exact weights of the chi-bar-square distribution and hence the exact $p$-value. Moreover, since the weights have  simple analytical expressions, the associated standard deviations are null.
Using the \pkg{EnvStats}, it is possible to print the results of the test using function \fct{print.htest}.

\begin{CodeChunk}
\begin{CodeInput}
R>  library("EnvStats")
R>  print.htest(vt)
\end{CodeInput}
\begin{CodeOutput}
Results of Hypothesis Test
--------------------------

Null Hypothesis:                 variance of age = 0

Alternative Hypothesis:          variance of age > 0

Test Name:                       Likelihood ratio test for variance components in mixed effects models

Data:                            

Test Statistic:                  LRT = 0.8326426

Test Statistic Parameters:       df        = 1, 2
                                 weights   = 0.5, 0.5
                                 sdweights = 0, 0
                                 FIM       = NA

P-values:                        pvalue.weights    = 0.5104889
                                 pvalue.sample     =        NA
                                 pvalue.lowerbound = 0.5104889
                                 pvalue.upperbound = 0.5104889
\end{CodeOutput}
\end{CodeChunk}

If we re-run the function with the option \code{pval.comp="both"}, we get the same results since the weights are explicit in this example. This time, we run the function with fits from \pkg{nlme} to show that results are similar if not identical (differences can appear in the LRT statistic due to the way the log-likelihood is computed within each package). 

\begin{CodeChunk}
\begin{CodeInput}
R>  vt <- varCompTest(lm1.h1.nlme, lm1.h0.nlme, pval.comp = "both")
R>  summary(vt)
\end{CodeInput}
\begin{CodeOutput}
Variance components testing in mixed effects models
Testing that variance of  age is null

 Likelihood ratio test statistic:
	LRT =  0.8331072

 Limiting distribution:
	mixture of 2 chi-bar-square distributions with degrees of freedom 1 2
	associated weights (and sd): 0.5 (0) 0.5 (0)

 p-value of the test:
	from estimated weights: 0.5103454
	bounds on p-value: lower  0.5103454 upper  0.5103454
\end{CodeOutput}
\end{CodeChunk}

\textsc{Case 2:} testing that the variance of age is null in a model with independent random effects. 
The number of components in the mixture is the same as in \textsc{Case 1}, but the degrees of freedom have been shifted downward. Note that the weights  have also simple analytical expressions in this case, as well as the $p$-value. Results are the same with the \pkg{nlme} package.

\begin{CodeChunk}
\begin{CodeInput}
R>  vt <- varCompTest(lm2.h1.lme4, lm2.h0.lme4, pval.comp = "both")
R>  summary(vt)
\end{CodeInput}
\begin{CodeOutput}
Variance components testing in mixed effects models
Testing that variance of  age is null

 Likelihood ratio test statistic:
	LRT =  0.5304106

 Limiting distribution:
	mixture of 2 chi-bar-square distributions with degrees of freedom 0 1
	associated weights (and sd): 0.5 (0) 0.5 (0)

 p-value of the test:
	from estimated weights: 0.2332171
	bounds on p-value: lower  0.2332171 upper  0.2332171
\end{CodeOutput}
\end{CodeChunk}	

\textsc{Case 3:} testing the presence of randomness in the model, i.e., testing that the variances of age and of the intercept are null, in a model with independent random effects.

\begin{CodeChunk}
\begin{CodeInput}
R>  vt <- varCompTest(lm3.h1.nlme, lm3.h0)
R>  summary(vt)
\end{CodeInput}
\begin{CodeOutput}
Variance components testing in mixed effects models
Testing that variances of Intercept and age are null

 Likelihood ratio test statistic:
	LRT =  50.13311

 Limiting distribution:
	mixture of 3 chi-bar-square distributions with degrees of freedom 0 1 2

 p-value of the test:
	bounds on p-value: lower  7.18311e-13 upper  7.215163e-12
\end{CodeOutput}
\end{CodeChunk}	

In this case, using the provided upper bound we see that the exact $p$-value is smaller than $10^{-11}$, which is enough in practice to reject the null hypothesis without computing the chi-bar-square weights and the associated $p$-value. It is also possible to compute the weights using the option \code{pval.comp="both"}. In this case, the weights are explicit but depend on the Fisher Information Matrix (FIM).
Since we are dealing with a linear mixed-effects model, the FIM can be extracted from \pkg{nlme} or \pkg{lme4} packages, using the option \code{fim="extract"}. This is the default behaviour of \code{varCompTest} function. Note that results can differ between packages since the methods used to compute the FIM is not the same in \pkg{nlme} and \pkg{lme4}. 

Using \pkg{nlme} package:

\begin{CodeChunk}
\begin{CodeInput}
R>  vt <- varCompTest(lm3.h1.nlme, lm3.h0, pval.comp = "both")
R>  summary(vt)
\end{CodeInput}
\begin{CodeOutput}
Variance components testing in mixed effects models
Testing that variances of Intercept and age are null

 Likelihood ratio test statistic:
	LRT =  50.13311

 Limiting distribution:
	mixture of 3 chi-bar-square distributions with degrees of freedom 0 1 2
	associated weights (and sd): 0.3765372 (0) 0.5000000 (0) 0.1234628 (0)

 p-value of the test:
	from estimated weights: 2.32255e-12
	bounds on p-value: lower  7.18311e-13 upper  7.215163e-12
\end{CodeOutput}
\end{CodeChunk}	

Using \pkg{lme4} package:

\begin{CodeChunk}
\begin{CodeInput}
R>  vt <- varCompTest(lm3.h1.lme4,lm3.h0.lme4, pval.comp = "both")
R>  summary(vt)
\end{CodeInput}
\begin{CodeOutput}
Variance components testing in mixed effects models
Testing that variances of (Intercept) and age are null

 Likelihood ratio test statistic:
	LRT =  50.13311

 Limiting distribution:
	mixture of 3 chi-bar-square distributions with degrees of freedom 0 1 2
	associated weights (and sd): 0.2500962 (0) 0.5000000 (0) 0.2499038 (0)

 p-value of the test:
	from estimated weights: 3.965487e-12
	bounds on p-value: lower  7.18311e-13 upper  7.215163e-12
\end{CodeOutput}
\end{CodeChunk}	

To compute the FIM using parametric bootstrap, one should use the option \code{fim="compute"}. The default bootstrap sample size is $B=1000$, but it can be changed using the \code{control} argument. When the FIM is computed via bootstrap, results are more consistent between \pkg{nlme} and \pkg{lme4}. 

Using \pkg{nlme} package:
\begin{CodeChunk}
\begin{CodeInput}
R>  vt <- varCompTest(lm3.h1.nlme, lm3.h0, pval.comp = "both", 
+     fim = "compute")
R>  summary(vt)
\end{CodeInput}
\begin{CodeOutput}
Variance components testing in mixed effects models
Testing that variances of Intercept and age are null

 Likelihood ratio test statistic:
	LRT =  50.13311

 Limiting distribution:
	mixture of 3 chi-bar-square distributions with degrees of freedom 0 1 2
	associated weights (and sd): 0.3463632 (0) 0.5000000 (0) 0.1536368 (0)

 p-value of the test:
	from estimated weights: 2.714622e-12
	bounds on p-value: lower  7.18311e-13 upper  7.215163e-12
\end{CodeOutput}
\end{CodeChunk}

Using \pkg{lme4} package:

\begin{CodeChunk}
\begin{CodeInput}
R> vt <- varCompTest(lm3.h1.lme4, lm3.h0, pval.comp = "both", 
+     fim = "compute")
R> summary(vt)
\end{CodeInput}
\begin{CodeOutput}
Variance components testing in mixed effects models
Testing that variances of (Intercept) and age are null

 Likelihood ratio test statistic:
	LRT =  50.13311

 Limiting distribution:
	mixture of 3 chi-bar-square distributions with degrees of freedom 0 1 2
	associated weights (and sd): 0.3562174 (0) 0.5000000 (0) 0.1437826 (0)

 p-value of the test:
	from estimated weights: 2.58658e-12
	bounds on p-value: lower  7.18311e-13 upper  7.215163e-12
\end{CodeOutput}
\end{CodeChunk}

\subsubsection{Generalized linear model}
Results from the test of the null hypothesis that there is no random effect associated with the herd are given by:

\begin{CodeChunk}
\begin{CodeInput}
R>  vt <- varCompTest(glm1, glm0)
R>  summary(vt)
\end{CodeInput}
\begin{CodeOutput}
Variance components testing in mixed effects models
Testing that variance of  (Intercept) is null

 Likelihood ratio test statistic:
	LRT =  14.00527

 Limiting distribution:
	mixture of 2 chi-bar-square distributions with degrees of freedom 0 1
	associated weights (and sd): 0.5 (0) 0.5 (0)

 p-value of the test:
	from estimated weights: 9.114967e-05
	bounds on p-value: lower  9.114967e-05 upper  9.114967e-05
\end{CodeOutput}
\end{CodeChunk}

\subsubsection{Nonlinear model}
As detailed in Section \ref{sec:weight}, the limiting distribution of the LRT for the test of the null hypothesis that only the asymptote is random against the alternative hypothesis of a full covariance matrix between the random effects is a mixture of 4 chi-square distributions with degrees of freedom varying between 2 and 5. When run with the default options, the \code{varCompTest} function gives:

\begin{CodeChunk}
\begin{CodeInput}
R>  vt <- varCompTest(nlm1.nlme, nlm0.nlme)
R>  summary(vt)
\end{CodeInput}
\end{CodeChunk}
\begin{CodeChunk}
\begin{CodeOutput}
Variance components testing in mixed effects models
Testing that variances of R0 and lrc are null

Likelihood ratio test statistic:
  LRT =  2.519869

Limiting distribution:
  mixture of 3 chi-bar-square distributions with degrees of freedom 0 1 2

p-value of the test:
  bounds on p-value: lower  0.05620995 upper  0.1980462
\end{CodeOutput}
\end{CodeChunk}

\begin{CodeChunk}
\begin{CodeInput}
R>  varCompTest(nlm1.lme4, nlm0.lme4)
R>  summary(vt)
\end{CodeInput}
\end{CodeChunk}
\begin{CodeChunk}
\begin{CodeOutput}
Variance components testing in mixed effects models
Testing that variances of R0 and lrc are null

 Likelihood ratio test statistic:
	LRT =  2.456656

 Limiting distribution:
	mixture of 3 chi-bar-square distributions with degrees of freedom 0 1 2

 p-value of the test:
	bounds on p-value: lower  0.05851384 upper  0.2049047
\end{CodeOutput}
\end{CodeChunk}

Depending on the desired level of the test, more precision may be needed on the $p$-value.
For models fitted with \pkg{nlme}, this can be obtained using the options \code{pval.comp = 'both'} or \code{pval.comp = 'approx'} and \code{fim = 'extract'} to use the FIM computed by \code{nlme}, or \code{fim = 'compute'} to approximate the FIM using parametric bootstrap, with a default bootstrap sample size of 1000.

\begin{CodeChunk}
\begin{CodeInput}
R>  vt <- varCompTest(nlm1.nlme, nlm0.nlme, pval.comp = "both", 
+     fim = "compute")
R>  summary(vt)
\end{CodeInput}
\end{CodeChunk}
\begin{CodeChunk}
\begin{CodeOutput}
Variance components testing in mixed effects models
Testing that variances of R0 and lrc are null

 Likelihood ratio test statistic:
	LRT =  2.519869

 Limiting distribution:
	mixture of 3 chi-bar-square distributions with degrees of freedom 0 1 2
	associated weights (and sd): 0.2490444 (0) 0.5000000 (0) 0.2509556 (0)

 p-value of the test:
	from estimated weights: 0.1273992
	bounds on p-value: lower  0.05620995 upper  0.1980462
\end{CodeOutput}
\end{CodeChunk}

Here again, the weights are exact, once the FIM is known, so that the standard deviations for each weight is 0. For information purposes, the above code took approximatively 175 seconds on a laptop with a 8 cores Intel(R) Core(TM) i5-8250U CPU @ 1.60GHz processor.

For nonlinear models fitted with \pkg{lme4}, to the best of our knowledge no method is available to extract the FIM (although \pkg{merDeriv} provides the FIM for linear and generalized linear models fitted via \pkg{lme4}). Only the option \code{fim='compute'} can then be used.
We get the following results, which are very similar to the ones obtained with the \pkg{nlme} package. However the code was a bit longer to run, and took 430 seconds.

\begin{CodeChunk}
\begin{CodeInput}
R > varCompTest(nlm1.lme4, nlm0.lme4, pval.comp = "both", 
+     fim = "compute")
\end{CodeInput}
\end{CodeChunk}
\begin{CodeChunk}
\begin{CodeOutput}
Variance components testing in mixed effects models
Testing that variances of R0 and lrc are null

 Likelihood ratio test statistic:
	LRT =  2.456656

 Limiting distribution:
	mixture of 3 chi-bar-square distributions with degrees of freedom 0 1 2
	associated weights (and sd): 0.2590517 (0) 0.5000000 (0) 0.2409483 (0)

 p-value of the test:
	from estimated weights: 0.1290591
	bounds on p-value: lower  0.05851384 upper  0.2049047
\end{CodeOutput}
\end{CodeChunk}

\subsection{Estimation of the FIM by parametric bootstrap}
To evaluate the performance of the parametric bootstrap estimation of the Fisher Information Matrix implemented in \pkg{varTestnlme}, we compared the empirical coverages of the nominal 95\% confidence intervals based on the asymptotic normality property of the MLE obtained using the FIM extracted from packages \pkg{nlme}, \pkg{lme4} and \pkg{saemix}, and our estimate.

First, a linear mixed-effects model with two correlated random effects was fitted to $R=1000$ simulated datasets of size $n=100$.
\begin{align*}
	y_{ij} & = \beta_0 + \beta_1 t_{j} + b_{i0} + b_{i1} t_j + \varepsilon_{ij}, \quad i=1,\dots, n, j=1, \dots, 20\\
	\begin{pmatrix}
	b_{i0} \\
	b_{i1}
	\end{pmatrix} & \sim \mathcal{N}\left(\begin{pmatrix}
	0 \\0
	\end{pmatrix}, \begin{pmatrix}
	\gamma_1^2 & \gamma_{12} \\
	\gamma_{12} & \gamma_2^2
\end{pmatrix}	 \right), \quad \varepsilon_{ij} \sim \mathcal{N}(0,\sigma^2)
\end{align*}

We took $\beta_0 =5, \beta_1=7, \gamma_1 = 0.8, \gamma_2 = 1, \gamma_{12} = 0.4$ and $\sigma = 1.2$, with 
The FIM was extracted and estimated for each of these $R$ datasets, with a bootstrap sample size $B \in \{100, 1000\}$.
Results are given in Table \ref{tab:boot_lin}. Results show that, as expected, the bootstrap estimate is better when the bootstrap sample size is increased, but good results were already obtained for small values of $B$.  Interestingly, the bootstrap estimate lead to better coverages than the package estimate when using \pkg{lme4}, especially for variance components. This is crucial since the FIM components involved in the chi-bar-square weights computation are those corresponding to the variance components.

We performed the same comparisons on a nonlinear model similar as the one presented in \eqref{eq:nonlinmod}, with three correlated random effects. We took $\beta = (200,725,150)^\top$, for the variances we took respectively $20,72.5,15$ (i.e., 10\% of the mean values), and for the correlations we took $0.9$ between $Asym_i$ and $R_{0,i}$, and 0.1 between all the other parameters, and $\sigma = 10$.. Results are presented in Table and \ref{tab:boot_nonlin_nondiag}. Recall that \pkg{lme4} does not allow to estimate the FIM in nonlinear mixed effects models.  The added value of the bootstrap estimation of the FIM becomes obvious for nonlinear mixed-effects models with correlated random effects. Indeed, in this case the FIM components corresponding to the variance parameters are not well estimated by the default function of \pkg{nlme} package, leading to particular bad empirical coverages. Results are better when the bootstrap estimate of the FIM is performed using parameter estimates obtained with \pkg{lme4} than those obtained with \pkg{nlme}.

\begin{table}
\centering
\caption{Empirical coverage of the 95\% asymptotic confidence intervals according to the type of FIM estimate, in a linear mixed-effect model with two correlated random effects, computed on 1000 repetitions.}
\label{tab:boot_lin}
\begin{tabular}{c|cc|cc|cc}
  \hline
\multirow{3}{*}{Parameter}& \multicolumn{2}{c|}{Extracted} &  \multicolumn{4}{c}{Boostrap} \\
\cline{2-7}
& \multirow{2}{*}{\pkg{nlme}} & \multirow{2}{*}{\pkg{lme4}} & \multicolumn{2}{c|}{using \pkg{nlme} estimates} & \multicolumn{2}{c}{using \pkg{lme4} estimates} \\  
\cline{4-7}
& & & $B=100$ & $B=1000$ & $B=100$ & $B=1000$ \\
  \hline
$\beta_1$ & 0.948 & 0.956 & 0.948 & 0.953 & 0.948 & 0.945 \\ 
$\beta_2$ & 0.935 & 0.944 & 0.942 & 0.929 & 0.932 & 0.932 \\ 
$\gamma_1^2$ & 0.937 & 0.928 & 0.936 & 0.966 & 0.922 & 0.940 \\ 
$\gamma_2^2$ & 0.879 & 0.938 & 0.938 & 0.935 & 0.938 & 0.933 \\ 
$\gamma_{1,2}$ & 0.979 & 0.943 & 0.947 & 0.946 & 0.941 & 0.949 \\ 
$\sigma^2$ & 0.956 & 1 & 0.950 & 0.957 & 0.938 & 0.956 \\ 
   \hline
\end{tabular}
\end{table}


\begin{table}
\centering
\caption{Empirical coverage of the 95\% asymptotic confidence intervals according to the type of FIM estimate, in a nonlinear mixed-effect model with three correlated random effects, computed on 1000 repetitions.}
\label{tab:boot_nonlin_nondiag}
\begin{tabular}{c|c|cc|cc}
  \hline
\multirow{3}{*}{Parameter}& Extracted &  \multicolumn{4}{c}{Boostrap} \\
\cline{2-6}
& \multirow{2}{*}{\pkg{nlme}} & \multicolumn{2}{c|}{using \pkg{nlme} estimates} & \multicolumn{2}{c}{using \pkg{lme4} estimates} \\  
\cline{3-6}
& &$B=100$ & $B=1000$ & $B=100$ & $B=1000$ \\
  \hline
$\beta_1$ & 0.934 &  1& 1 & 0.920 & 0.923\\ 
$\beta_2$ & 0.897 &  1 & 1  & 0.942 & 0.953\\  
$\beta_3$ & 0.945 &  0.983 & 0.977 & 0.923 & 0.923\\  
$\gamma_1^2$ & 0.926 & 1& 1 & 0.899 & 0.911\\ 
$\gamma_2^2$ &0.982 & 1 & 1 & 0.959 &0.989\\ 
$\gamma_3^2$ & 1 & 0.994 & 0.994 & 0.901 &0.885\\ 
$\gamma_{1,2}$ & 0.026 & 0.898 & 0.900 & 0.957 & 0.992\\
$\gamma_{1,3}$ & 0.006 & 0.479 & 0.467 & 0.895 &0.915\\ 
$\gamma_{2,3}$ &0.134 & 0.999 & 1 & 0.963 &0.998\\ 
$\sigma^2$ & 0.945 & 1 & 1 &0.968 &0.975\\ 
   \hline
\end{tabular}
\end{table}


\section{Summary and discussion} \label{sec:summary}
In this paper, we present the \proglang{R} package \pkg{varTestnlme}, which provides a unified framework for variance components testing in linear, generalized linear and nonlinear mixed-effects models fitted with \pkg{nlme}, \pkg{lme4} and \pkg{saemix}. The main function, \code{varCompTest}, performs the comparison of two nested models \code{m1} and \code{m0}. It is thus possible to test that the variances of any subset of random effects is null, taking into account correlation between them. It is also possible to test that one or several correlations are null, and to test simultaneously if some fixed effects are null. The tests are performed using likelihood ratio test. Additional  tools are provided to approximate the Fisher Information Matrix when it is needed to compute the limiting distribution of the likelihood ratio test statistic. 

Possible future developments of the package include in particular the consideration of more than one level of random effects, to account for multiple nested random effects. Another perspective would be to allow for more general structures for the covariance matrix $\Gamma$ of the random effects. This would encompass for example parametric structures such as compound symmetry, auto-regressive or with spatial correlation, as proposed in the \pkg{nlme} package.

%
%
%
%
%
%


\bibliography{refs}



\end{document}